\documentclass{iopart}
\usepackage[T1]{fontenc}
\usepackage[latin1]{inputenc}
\usepackage{bm}
\usepackage{amssymb}
\usepackage{mathrsfs}
\usepackage{latexsym}
\usepackage{amsfonts}
\usepackage{epsfig}
\usepackage{psfrag}

\newcommand{\nn}{\nonumber\\}

\newcommand{\f}[1]{\mbox{\boldmath$#1$}}

\newcommand{\na}{\mbox{\boldmath$\nabla$}}
\newcommand{\bea}{\begin{eqnarray}}
\newcommand{\ea}{\end{eqnarray}}
\newcommand{\bmult}{\begin{multline}}
\newcommand{\emult}{\end{multline}}
\newcommand{\eea}{\end{eqnarray}}
\newcommand{\ord}{{\cal O}}
\newcommand{\eqref}[1]{(\ref{#1})}
\newcommand{\bc}{\begin{center}}
\newcommand{\ec}{\end{center}}

\usepackage{color}
\newcommand{\col}[1]{\textcolor{red}{#1}}

\begin{document}

\title[Scaling of topological defect creation]{System size scaling 
of topological defect creation in a second-order dynamical 
quantum phase transition}

\author{Michael Uhlmann$^{1}$, Ralf Sch\"utzhold$^{2}$, and
Uwe R.~Fischer$^{3}$}

\address{
$^1$Department of Physics and Astronomy, University of British Columbia,
Vancouver, British Columbia, V6T 1Z1 Canada
\\
$^2$Fachbereich Physik, Universit\"at Duisburg-Essen, D-47048
   Duisburg, Germany \\
$^3$Center for
Theoretical Physics,  Department of Physics and Astronomy, 
Seoul National University, Seoul 151-747, Korea}

\begin{abstract}
We investigate the system size scaling of the net defect number
created by a rapid quench in a second-order quantum phase transition
from an $O(N)$ symmetric state to a phase of broken symmetry. 
Using a controlled mean-field expansion for large $N$, we find 
that the net defect number variance in convex volumina
scales like the surface area of the sample for short-range correlations.
This behaviour follows generally from spatial and internal symmetries. 
Conversely, if spatial isotropy is broken, e.g., by a lattice, and
in addition long-range periodic correlations develop in the
broken-symmetry phase, we get the rather counterintuitive result that
the scaling strongly depends on the dimension being even or odd:
For even dimensions, the net defect number variance 
scales like the surface area squared, with a prefactor 
oscillating with the system size, while for odd dimensions, 
it essentially vanishes. 
\end{abstract}


\maketitle

\maketitle

\section{Introduction}
Rapidly sweeping  through a second-order (quantum or thermal) 
{symmetry-breaking} phase transition in general {may trigger} 
the creation of topological
defects. This is due to the fact that the system cannot adjust abruptly to the
new order, so that (time-dependent) correlations are spontaneously
generated in the interacting many-body system. Correctly counting
the topological defects generated by the sweep amounts to an analysis
of these correlations.

The fundamental global requirement for topological defect generation is that
the symmetry group of the new phase permits the defects in the sense
that the homotopy group of the coset space, formed by the quotient
space of unbroken (old) and broken (new) symmetries is nontrivial.
The physical consequences of the latter mathematical fact were first
used by Kibble
\cite{Kibble} in the high-energy context of symmetry breakings from a
(grand unified) symmetry group of generally high dimension to one of
lower dimension in the earliest stages of the universe. This led to
the {classification} 
of the symmetry group(s) for which cosmic strings and
domain walls can, in
principle, be produced by the universe's expansion after the big bang.
An implementation of this idea   
directly accessible to laboratory-scale experimental realizations in condensed-matter systems
was devised nine years later by Zurek \cite{Zurek}. Hence the 
historically developed and by now established name {\em Kibble-Zurek mechanism} 
for this  particular symmetry-breaking based scenario of topological defect formation.
While Kibble realized that in the early universe relativistic causality alone mandates the appearance of defects according to the nontrivial homotopy groups
relevant to the symmetry-breaking scheme at hand, Zurek  
used scalings of the relaxation time and healing length with the quench time
in the vicinity of the critical point to predict topological defect densities. The
scaling behaviors derived by Zurek, based on the critical exponents of the phase transition under consideration, therefore led to concretely testable experimental signatures of the 
general idea of Kibble. 


\bigskip
\begin{figure}[!bh]
\bc
\includegraphics[angle=0,width=0.45\textwidth]{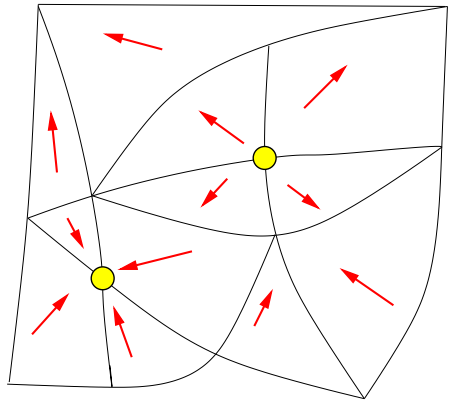}
\caption{\label{KZ}
    Origin of topological defect formation due to the Kibble-Zurek
    mechanism:
    Rapidly sweeping through a second-order phase transition
    statistically generates domains due to the finite propagation
    speed communicating of the new order in space and establishing the
    corresponding order parameter correlations.
    If by chance, at a given vertex, the domains meet such that a
    (generalized) "phase winding" is produced, a topological defect with
    a core of the old symmetry (yellow) is formed.
    In the simple example shown in the figure, a $U(1)$ order
    parameter in the two-dimensional plane with field direction
    indicated by the red arrows, this amounts to the order parameter
    vector orientation winding around the core by $2\pi$, for the planar spin hedgehog            depicted.
}
\ec
\end{figure}

The basic physical phenomenon is depicted in Fig.\,\ref{KZ} for the
simple case of a $U(1)$ order parameter in the plane:
It relies on the finite propagation speed
communicating the new order and the corresponding correlations
between spatially well-separated regions.
These finite communication speeds lead to the whole sample not being
able to establish the new order homogeneously throughout, thus causing
an inhomogeneous order parameter distribution.
Order parameter domains form, which, with some appropriate statistical
probability, potentially enclose topological defects.

The Kibble-Zurek mechanism {may occur} in many diverse physical 
settings, for
instance, in nonequilibrium phase transitions during the early universe
\cite{Boyanovsky,Stephens}, and has been measured in various condensed
matter systems like liquid crystals \cite{Bowick} or superconducting films
\cite{Maniv}.
There has been a tremendously increasing interest in the closely
related nonequilibrium quantum many-body phenomena during 
{quantum} phase transitions in recent years, cf.,  e.g.,
\cite{ZDZ,Damski,Dziarmaga,PRL,BHM,Klich,DamskiZurek07,Lamacraft,
Saito,Cherng,Altland,Polkovnikov,Gritsev,Sen,Dutta,DziarmagaReview},
due also to the fact that for
controlled dynamical entanglement
generation \cite{Cubitt} the dynamical evolution of quantum many-body
correlations is strongly relevant.

A clean experimental evidence for Kibble-Zurek mediated defect
formation in dilute cold atomic gases, which are an experimentally
precisely {\it in situ} controllable variant of conventional condensed
matter systems, has been obtained in spinor as well as
scalar Bose gases \cite{Sadler,Weiler}.
These dilute gases are particularly suitable to test predictions of the
dynamical Kibble-Zurek theory because of their comparatively long
re-equilibration times, which effectively permit a real-time study of the
evolution of perturbations in the spatial many-body correlations and thus
ultimately also a study of the creation process of topological
defects.
Furthermore, it was shown that the Kibble-Zurek type scaling of the defect number with the sweep rate through the phase transition depends on the
inhomogenity of the sample \cite{Laguna,Zurek09} (is becoming more strongly
dependent on the sweep rate), which is relevant for the intrinsically
inhomogenous harmonically trapped gases.
While most investigations to date have been carried out for bosonic systems in
low dimension (where {\em dimension} here refers to real as well as
order-parameter dimension), due to the rapidly growing  interest in cold atomic gases, more recently fermionic problems in these systems have also been tackled,
e.g. in  \cite{Moeckel,Babadi}.

\begin{figure}[!t]
  \bc
  \includegraphics[angle=0,width=0.6\textwidth]{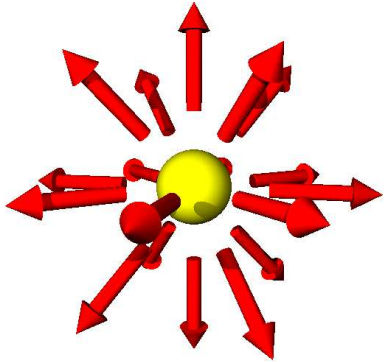}
  \caption{\label{illustration}
    In high dimensions, a rapidly increasing degree of coordination of
    the order parameter orientation in neighboring domains is
    necessary to enclose a defect.
    In the three-dimensional example shown above for the classic
    monopole configuration, all arrows in the surrounding domains
    (red) have to point away from the central defect core (yellow),
    cf.\,\,the two-dimensional situation in Fig.\,\ref{KZ} which
    already requires significantly less coordination.
  }
  \ec
\end{figure}

When calculating defect densities from correlation functions,
{frequently} 
the excitation densities (quasiparticle densities) and the topological 
defect densities actually created by the quench are assumed to be in a 
one-to-one correspondence cf., e.g., \cite{Polkovnikov,Sen,Dutta}.
This identification is however misleading in general, because the necessary
correlations created by the quench are much stronger for creating
topological defects as compared to the correlations for
quasiparticle excitations in high dimension
\cite{PRD}.
For an illustration of this fact cf.\,Fig.\ref{illustration}. 
A topological defect at a given vertex enclosed by domains after the
quench implies that the direction vectors of all neighbouring domains
either point away or towards the vertex (for a monopole configuration 
of the spin vectors, see also below). 
{In a high-dimensional space, for example,} 
this condition for a defect core to be enclosed by the domains at the
vertex such that a winding of the order parameter orientation is
induced becomes increasingly difficult to fulfill, given that the
direction vectors orient themselves statistically inside the sample
(i.e. due to quantum or thermal fluctuations establishing the order
parameter correlations). 

In the following, we will investigate topological defect creation in
quantum quenches concentrating on the behaviour of the defect creation
rate for high internal and spatial dimension of the system.
To this end, we derive essentially exact statements on topological
defect production in the large $N$ limit of an $O(N)$ symmetric model
in $D=N$ spatial dimensions.
Similar to large-$N$ approaches in condensed matter and field theory
we argue that our predictions for the defect production will remain
qualitatively valid also for smaller dimensions.
More specifically, assuming that there is no critical value of $N$
where the system behaviour changes in a discontinous manner, we expect
our results 
to be qualitatively applicable also to finite $N$, e.g., $N=3$, 
which are accessible to experimental tests, for example in spinor
Bose-Einstein condensates with large spin, cf.\ the review
\cite{UedaKawaguchi}.

In the next Section, we begin by deriving a large $N$ expansion for a
general $O(N)$ model resulting in a bilinear effective action,
which is used to obtain the equations of motion describing
the instability of direction modes during the phase transition. 
We then apply these equations to calculate the behavior of the
spatial direction correlation functions {shortly} after the
transition. 
The latter result is used to deduce {\em ab initio} the winding
number variance for our $O(N)$ model. 
Finally, we delineate the behavior of the winding number variance 
with the size of the sample in dependence on the spatial behavior 
of the direction correlations (which is, e.g., short-range or
long-range periodic).

\section{Large-$N$ Expansion}

In order to illustrate the main idea -- i.e., how to obtain an effective
quadratic action for {a} 
$N$-component field
{$\bm\Phi = (\Phi_1,\Phi_2,\dots,\Phi_N)$} in the large $N$ limit,
let us consider the (Euclidean) generating functional of the non-linear
$O(N)$ model
     \bea \fl 
        Z_\beta[\f{J}]=
        \int\mathfrak{D}^N\f{\Phi}\;
        \exp\left\{-\frac12\int\limits_0^\beta d\tau
        \int d^Nr
        \left[
        \f{\dot\Phi}^2+
        V(\f{\Phi}^2)-
        c^2\f{\Phi}\cdot\na^2\f{\Phi}-
        2\f{J}\cdot\f{\Phi}
        \right]
        \right\}
        \,,
     \ea
with inverse temperature $\beta = 1/k_{\rm B}T$.
%
%
The {$O(N)$-invariant} potential $V$ assumes the form
     \bea
        V(\f{\Phi}^2)=m^2\f{\Phi}^2+\frac{g}{N}\,\f{\Phi}^4+\dots
        \,,
        \label{V_N}
     \ea
where the factor $1/N$ in the second term ensures a well-defined
large-$N$ limit.
{For $m^2>0$, we get an $O(N)$-symmetric ground state,  
$\langle\bm\Phi\rangle=0$, whereas $m^2<0$ induces symmetry breaking,  
$\langle\bm\Phi\rangle\neq0$.}
%
%
We can formally split the field into modulus $\chi \geq 0$ 
and direction $\bm n$ with $\bm n^2 = 1$
     \bea
        \bm \Phi(\bm r,\tau) \, = \,
        \chi(\bm r,\tau)\,\bm n(\bm r,\tau)
        \,.
     \ea
%
The functional integration measure then becomes
     \bea\fl 
        \mathfrak{D}^N\f{\Phi}=
        \|\mathcal S_{N-1}\|
        \,\mathfrak{D}\chi\,\chi^{N-1}\;
        \mathfrak{D}^{N-1}\f{n}
        \,,
        \label{Z_measure_spherical}
     \ea
where $\|\mathcal S_{N-1}\| = 2\pi^{N/2}/\Gamma(N/2)$ is the surface
area of the unit sphere in $N$ dimensions and thus we get for the  
generating functional
     \bea \fl 
        Z_\beta[\f{J}]=
        \|\mathcal S_{N-1}\|\int\mathfrak{D}^{N-1}\f{n}\;
        Z^\chi_\beta[\f{n},\f{J}]
     \ea
with the remaining integration over $\chi$
(note that $\f{n}\cdot\f{\dot n}=0$ by normalization)
     \bea \fl 
        Z^\chi_\beta[\f{n},\f{J}]=
        \int\mathfrak{D}\chi\;
        \exp\left\{
        \int\limits_0^\beta d\tau
        \int d^Nr
        \left(
        (N-1)\ln\chi-
        \frac12[\chi^2\f{\dot n}^2+\dot\chi^2+V(\chi^2)]
        +\dots
        \right)
        \right\}
        \,. \nn
        \label{Z_chi}
     \ea
In view of the phase-space factor $\chi^{N-1}$ in the measure
\eqref{Z_measure_spherical}, we get, for $N \gg 1$, a strong
enhancement of contributions for large $\chi$.
{On the other hand, the terms in the action in \eqref{Z_chi}
such as $\chi^2\f{\dot n}^2$ yield a strong suppression at large $\chi$.
Therefore, the $\chi$-integral will be dominated by the saddle point 
of the exponent in \eqref{Z_chi} given by}
%
%
     \bea
        \frac{N-1}{\chi}
        =
        \ddot\chi
        + \bm{\dot n}^2\chi
        + \frac12\frac{dV}{d\chi}
        + \dots
        \,.
     \ea
In the large $N$ limit, $\dot{\bm n}^2$ is the sum of many
independently fluctuating terms and thus approximately constant
(the same is true for $\bm n \cdot \na^2\bm n$).
Therefore, the saddle-point approximation yields a constant field
$\chi$, i.e., a classical (mean) field
     \bea \fl 
        \chi^2
        = \bm \Phi ^2
        \approx \chi^2_{\rm class}
        = \langle\bm \Phi^2 \rangle
        = \ord(N)
        \,.
     \ea
Due to the steep potential for the field $\chi$
(in the large $N$ limit), its quantum fluctuations can be neglected
and we can approximate it by a constant c-number.
{In the above formula, we have assumed a suitable ultraviolet
regularization resulting in finite expectation values
$\langle \bm\Phi^2\rangle = \ord(N)$
and $\langle \dot{\bm \Phi}^2\rangle = \ord(N)$.}

Consequently, to lowest order in $1/N$, we obtain the same results
as with a bilinear effective action 
     \bea \fl 
\label{bilinear}
        Z_\beta^{\rm eff}[\f{j}]=
        \int\mathfrak{D}^N\f{\phi}\;
        \exp\left\{-\frac12\int\limits_0^\beta d\tau
        \int d^Nr
        \left[
        \dot{\bm\phi}^2+
        m^2_{\rm eff}\bm{\phi}^2-
        c^2_{\rm eff}\bm{\phi}\cdot\na^2\bm{\phi}-
        2\bm{j}\cdot\bm{\phi}
        \right]
        \right\}
        \,,
     \ea
where the renormalized constants $m^2_{\rm eff}$ and $c^2_{\rm eff}$
depend on the shape of the nonlinear potential $V$.
Of course, this line of argument is very similar to the nonlinear
$\f{\sigma}$-model which does also approach a linear theory for $N\gg1$.
As one may easily imagine, the same procedure can be applied in more
general cases.
For example, one may include arbitrary dispersion relations by adding
terms like $\f{\phi}\cdot\na^4\f{\phi}$ and $\f{\phi}\cdot\na^6\f{\phi}$
etc.
Furthermore, the prefactors in front of the different terms
(such as $c^2$) may depend on $\f{\phi}^2$, similar to the potential  
$V(\f{\phi}^2)$.

The general idea is always the same:
a sum of many ($N\gg1$) independently fluctuating quantities on an
equal footing such as $\f{\phi}^2$ is approximately a c-number whose
fluctuations are suppressed by $1/\sqrt{N}$
(cf.~the de~Finetti theorem and the law of large numbers).
Therefore, we shall use the same linear effective action also in the
time-dependent case (with real time $t$ instead of Euclidean time
$\tau$) -- and with a possibly time-dependent
$\langle\f{\phi}^2\rangle(t)$, which is due to the growth of the
thermal and quantum fluctuations (see below).


\section{Phase transition and correlation function}

Having motivated an effectively quadratic action in the large $N$
limit, we can discuss the behaviour of the fluctuations during the
phase transition, by using a linear evolution equation of the form
     \bea
        F(-\na^2) \ddot{\bm \phi}
        + G(-\na^2) \bm \phi
        = 0 \,,
        \label{linFieldEq}
     \ea
where we assumed time-independent coefficients $F(-\na^2)$ and $G(-\na^2)$
{initially and after} the quench.
Note, however, that the validity of such linear equations is not
restricted to large $N$ but a linearization is often also possible in
lower dimensions, $N = \ord(1)$, using different expansion parameters
instead of $1/N$.
For instance, for a spinor Bose Einstein condensate,
the {\em gas parameter}, measuring the diluteness of the system, 
can be employed, cf.\ \cite{PRL}.

Before the quench, the field is in a stable configuration, i.e.,
all excitations have real frequencies.
The field operator can be written in terms of initial modes
      \bea\fl \qquad
      \hat{\bm \phi}(\bm r,t < t_{\rm in}) \, = \,
      \sum_\lambda     \int \frac{d^Dk}{(2\pi)^{D/2}}\,A_k
      \bm\varepsilon_{\bm k,\lambda}
      \left\{
      e^{i \bm k \bm r} e^{-i \omega_{k,<} t} \hat a_{\bm k,\lambda}
      + e^{-i \bm k \bm r} e^{i \omega_{k,<} t} \hat a_{\bm k,\lambda}^\dagger
      \right\} 
      \label{phi_in}
      \ea
with $N$ orthogonal polarizations,
$\bm \varepsilon_{\bm k, \lambda} \cdot \bm \varepsilon_{\bm k,\lambda'}
= \delta_{\lambda\lambda'}$.
The coefficients $A_k$ depend on the functions 
$F(k^2)$ and $G(k^2)$ in Eq.\,\eqref{linFieldEq} 
and are chosen such that the mode operators 
$\hat a_{\bm  k,\lambda}^\dagger$ and $\hat a_{\bm k,\lambda}$ creating
or annihilating a quasi-particle of wavenumber $\bm k$ and polarization
$\lambda$ obey the usual bosonic commutation relation
$[\hat a_{\bm k,\lambda},\hat a_{\bm k',\lambda'}^\dagger]
= \delta_{\bm k\bm k'}\delta_{\lambda\lambda}'$ for a bosonic field
$\hat{\bm \phi}$.
(In the following, we will take the basis vectors $\bm e_a$ labelling
the field components as the polarizations $\bm \varepsilon_{\bm k,\lambda}$.)

After quenching the system through a (second-order) phase transition,
{the $O(N)$-symmetric state is no longer the ground state of the system. 
As a result, some modes become unstable and thus}
the (instantaneous) dispersion relation {from \eqref{bilinear}}
dips below zero for some
wavenumbers, $\omega_{k,>}^2 < 0$, 
cf.\,Fig.\,\ref{dispersion}.
The time-dependence of $\hat{\bm\phi}$ can be expressed in terms
of instantaneous frequencies
      \bea\fl \qquad
      \hat{\bm \phi}(\bm r, t > t_{\rm out} )
      \, = \,
      \int \frac{d^N k}{(2\pi)^{N/2}} e^{i \bm k \bm r}
      \left[
        B_k e^{-i \omega_{\rm out} t} + C_k e^{i \omega_{\rm out} t}
      \right] \bm\varepsilon_{\bm k,\lambda} \hat a_{\bm k,\lambda}
      + {\rm H.c.}
      \label{phi_quench}
      \ea
with initial mode operators $\hat a_{\bm k,\lambda}$ and complex
coefficients $B_k$ and $C_k$ depending on the shape of the
transition.
Note that {many} 
of the modes are still stable after the quench and can be viewed as 
``virtual'' short-lived defect-antidefect pairs annihilating 
immediately after their creation.
These stable modes form a time-independent contribution to the
correlation function.
The field linearization outlined in the previous section (in the
large $N$ limit) is still valid for a finite time after the quench -- 
as long as the relative contribution of unstable modes is small
compared to the mean field $\langle\bm \phi^2 \rangle=\ord(N)$.

\begin{center}
\vspace*{1em}
\begin{figure}[b]
\psfrag{omega}{\large $\omega^2$}
\psfrag{k*}{\large $k_*$}
\psfrag{k}{\large $k$}
\centerline{\epsfig{file=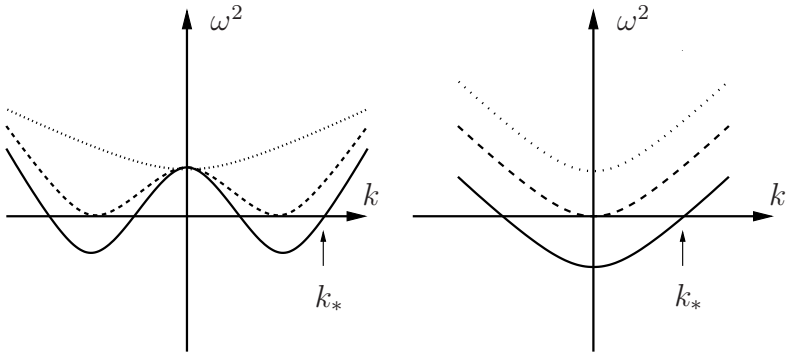,width=0.8\textwidth}}
\caption{\label{dispersion}
Two generic examples for the evolution of the dispersion relation
following from Eq.\,\eqref{linFieldEq} during a symmetry-breaking phase transition.
%
Initially (dotted line), all $k$-values are stable, $\omega^2(k)>0$.
At the critical point (dashed line), the dispersion relation
touches the $k$-axis, and after the transition (solid line),
modes in a finite $k$-interval become unstable, $\omega^2(k)<0$.
The left panel corresponds to a case, where $\omega^2(k=0)=m^2$
in Eq.\,\eqref{dispersion} remains positive while $c^2$ changes
sign; 
while in the right panel, $\omega^2(k=0)=m^2$ becomes negative.
In both cases, however, there is a dominant wave vector $k_*$
(for large $N$),
as indicated by the vertical arrows.}
\end{figure}
\vspace*{-1em}
\end{center} 

Bearing in mind that we start in the $O(N)$ symmetric phase,
which has $\langle\hat{\bm\phi}\rangle = 0$, we cannot simply
insert
$\bm n = \langle \hat{\bm \phi}\rangle/|\langle\hat{\bm\phi}\rangle|$
into the general expression for the winding number 
{given by Eq.\,\eqref{windop} below}.
Instead, in order to address the winding number statistics, we
must employ some appropriately defined direction operator $\hat{\bm n}$,
see \cite{PRD}.
Since the new phase grows from unstable fluctuations, the basic domain
structure with broken $O(N)$ symmetry and thus also the topological
defects {are} seeded by the field distribution of the unstable 
modes shortly after the quench.
Stable excitations, on the other hand, can be interpreted as virtual
defect-antidefect pairs coming only briefly into existence.
Especially for large $k$, these short-lived virtual defect-antidefect 
pairs would dominate the defect number in a given volume for any fixed time 
due to the phase space factor $k^{N-1}$ stemming from the $d^Nk$-integral.
This corresponds to the strong singularity of the two-point function 
$1/|\bm r-\bm r'|^{N-1}$ in $N\gg1$ dimensions. 
Since we are not interested in these short-lived pairs, but rather
in real long-lived defects originating from unstable fluctuations,
we introduce a time-averaged direction operator
      \bea 
      \hat{\bm n}(\bm r)  = 
      \frac{1}{\hat Z} \int dt\, g(t) \hat{\bm \phi}(\bm r,t) , \label{smoothing}
\ea
with (temporal) regularization function $g(t)$.


This time average supresses all short-lived pairs at large $k$.
Since the spatial behaviour of the $\f{n}$-correlator (\ref{hatnfull}) will 
be dominated by one wavenumber $k_*$, see Eq.~(\ref{corr}) below, and the 
time-dependence cancels due to the normalization of $\f{n}$, our result is 
independent of the choice of $g(t)$.
The only conditions on $g(t)$ are that its support lies well within the region 
of applicability of the effective action (\ref{bilinear}) and that $g(t)$ is 
smooth enough.
More precisely, the spectral width of $\tilde g(\omega)$ determines the 
uncertainty of $k=k_*\pm\Delta k$ which should be small $\Delta k\ll k^*$. 

Via the same arguments as in the previous Section, the normalization
factor $\hat Z=\ord(N)$ can be approximated by a c-number for large $N$ 
\bea     
      \hat Z^2  = 
        \left[
        \int dt g(t) \hat{\bm \phi}(\bm r,t)
        \right]^2
        \, = \,
        \left\langle
        \left[\int dt g(t) \hat{\bm \phi}(\bm r,t)
        \right]^2
        \right\rangle
        + \ord(N) .
      \label{n_def}
      \ea
Obviously, the rapidly oscillating parts of $\hat{\bm \phi}$, 
i.e., stable modes, are smoothed out in Eq.\,\eqref{smoothing} 
and only the exponentially growing excitations remain.
Note that the leading order of the normalization factor, $\hat Z\approx Z$,
would yield exact normalization of the expectation value,
$\langle \hat{\bm n}^2\rangle = 1$, but only approximate normalization
of the operator itself, $\hat{\bm n}^2 \approx 1$.
Proper normalization of the direction operator, $\hat{\bm n}^2 = 1$,
can only be achieved by including operator-valued higher orders of
$\hat Z$ as well.
In view of the mode expansion \eqref{phi_quench}, we get for
the time-averaged direction operator
      \bea
      \hat{\bm n}
      \, = \,
      \frac{1}{\hat Z} \int\limits_{\rm unstable}
      \frac{d^Nk}{(2\pi)^{N/2}}
      e^{i \bm k \bm r} \tilde C_k
      \bm \varepsilon_{\bm k, \lambda} \hat a_{\bm k,\lambda}
      + {\rm H.c.} \,,\label{average_n}
      \ea
where the integration domain in $k$ space runs over all unstable
modes.
The coefficients $\tilde C_k$ include the above time-averaging
{and the exponential growth of the unstable modes}, as well as  
the original coefficients $B_k$ and $C_k$, and possibly the initial 
distribution, e.g., of thermal origin,
of the occupation numbers of the different modes.

The calculation of the correlation function is now straightforward.
Taking the basis vectors $\bm e_a$ labelling the components $\phi_a$
of the field as the polarizations $\bm\varepsilon_{\bm k,\lambda}$ and
approximating the normalization factor by its expectation value,
we obtain
     \bea
        \left\langle
        \hat n_a(\bm r) \hat n_b(\bm r')
        \right\rangle
        \, = \,
        \frac{\delta_{ab}}{Z^2}
        \int \frac{d^Nk}{(2\pi)^N}
        e^{i\bm k(\bm r - \bm r')}
        |\tilde C_k|^2 \,. \label{hatnfull}
     \ea
In the case of spatial homogenity and isotropy (as opposed to internal
isotropy {of} the $N$ fields $\phi_a$) after the quench, where no
direction is distinguished, we can adopt spherical
coordinates and carry out the angular integration
      \bea 
      \left\langle
      \hat n_a(\bm r) \hat n_b(\bm r')
      \right\rangle
      \, = \,
      \frac{\delta_{ab}}{Z^2} \frac{1}{(2\pi)^{N/2}}
      \int dk k^{N-1} |\tilde C_k|^2
      \frac{J_\nu(k|\bm r - \bm r'|)}{(k|\bm r - \bm r'|)^\nu} \,,
      \label{nnkint}
      \ea
where {the index $\nu$ of the Bessel function $J_\nu$ is given by}
$\nu = N/2 - 1$.
The remaining $k$ integral is often dominated by one wavenumber
$k_*$:
In large dimensions $N \gg 1$, the phase-space factor $k^{N-1}$
strongly favors contributions near the largest unstable wavenumber,
cf.\ Fig.\,\ref{dispersion},
whereas for small $N = \ord(1)$, the dominant wavenumber is usually
determined by the largest coefficient $\tilde C_k$, i.e., the
fastest-growing mode.
In either case, as a result of the existence of a dominant wavenumber
$k_*$ and the ensuing application of the saddle-point integration, the
correlation function of the time-averaged direction vector now reads
      \bea
      \left\langle
      \hat n_a(\bm r) \hat n_b(\bm r')
      \right\rangle
      \, = \,
      \delta_{ab} \frac{2^\nu \Gamma(\nu +1 )}{N}
      \frac{J_\nu(k_*|\bm r - \bm r'|)}{{(k_*|\bm r - \bm r'|)^\nu}} \,.
      \label{corr}
      \ea
Note again that this is only a part of the direction correlations,  
generated by the unstable modes, and thus responsible for the formation of 
topological defects.
All stable modes have been averaged out in definition \eqref{n_def}
of the direction operator $\hat{\bm n}$.

For large $N \to \infty$, we can use the well-known asymptotic
behavior of Bessel functions \cite{Abramowitz} and
obtain Gaussian correlations
      \bea
      \left\langle
      \hat n_a (\bm r) \hat n_b (\bm r')
      \right\rangle
      \, = \,
      \frac{\delta_{ab}}{N}
      \exp\left\{ - \frac{{k_*^2 (\bm r - \bm r')^2}}{2N}
      \right\}\,,
      \label{Gaussian}
      \ea
which fall off on a typical distance $\sqrt{N}/k_*$.
Note that, since the corrections to the normalization factor $\hat Z$
of the direction operator are suppressed for large $N$, cf.\ \eqref{n_def},
these Gaussian correlations represent an exact result.
The typical oscillatory behavior of Bessel functions is exponentially
suppressed for large $N \to \infty$.
For finite $N$, the spatial decay of the correlations is too weak
such that the oscillations are retained.
However, for finite $N$, one should also be aware that corrections to
the normalization $\hat Z$ would become important, though the leading
order of the directional correlations would still be given by expression
\eqref{corr}.

\section{Topological Defects}

\subsection{Winding numbers in dimension $N$: Classification of topological defects}
\label{SecDefects}

As a quantum version of the classical winding number 
{counting the net number of topological defects within a domain 
$\mathcal M$} in $N$ dimensions \cite{Abanov}, we introduce 
a winding number operator $\hat{\mathfrak N}$ for the direction 
operator $\hat{\bm n}$ of the $O(N)$ field, cf.\ \cite{PRD}
      \bea
        \hat{\mathfrak N}(\mathcal M) =
        \frac{\varepsilon_{abc...}\varepsilon^{\alpha\beta\gamma...}}
                {\Gamma(N) \|\mathcal S_{N-1}\|}
                \oint\limits_{\partial \mathcal M}
                        dS_\alpha \hat n^a
                        (\partial_\beta \hat n^b)
                        (\partial_\gamma \hat n^c)... \,,
        \label{windop}
      \ea
where $\| \mathcal S_{N-1}\| = 2 \pi^{N/2}/\Gamma(N/2)$ 
is the surface area of the $N$ dimensional unit sphere.
The operator $\hat{\mathfrak N}$ 
(and similarly the classical winding number) 
remains unchanged if we subject the fields $\bm n$ to a global 
internal $O(N)$ rotation which can be smoothly deformed to the 
identity transformation.
Improper rotations, which are also $O(N)$ symmetry transformations
(but cannot be smoothly deformed to the identity transformation),  
generally do not leave $\hat{\mathfrak N}$ invariant.

\begin{figure}[hbt]
  \begin{minipage}{0.245\textwidth}
    \includegraphics[width=\textwidth]{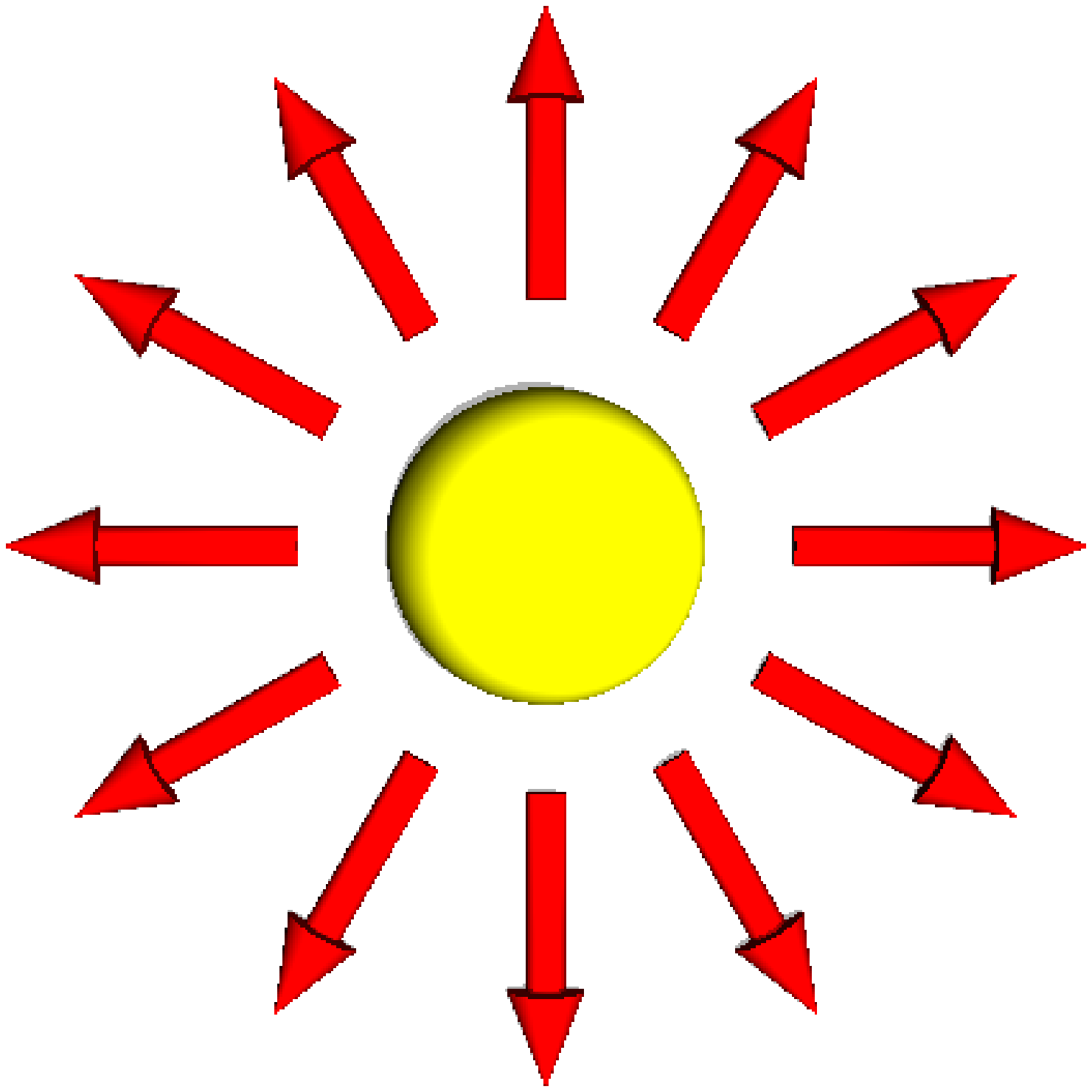}
  \end{minipage}
  \begin{minipage}{0.245\textwidth}
    \includegraphics[width=\textwidth]{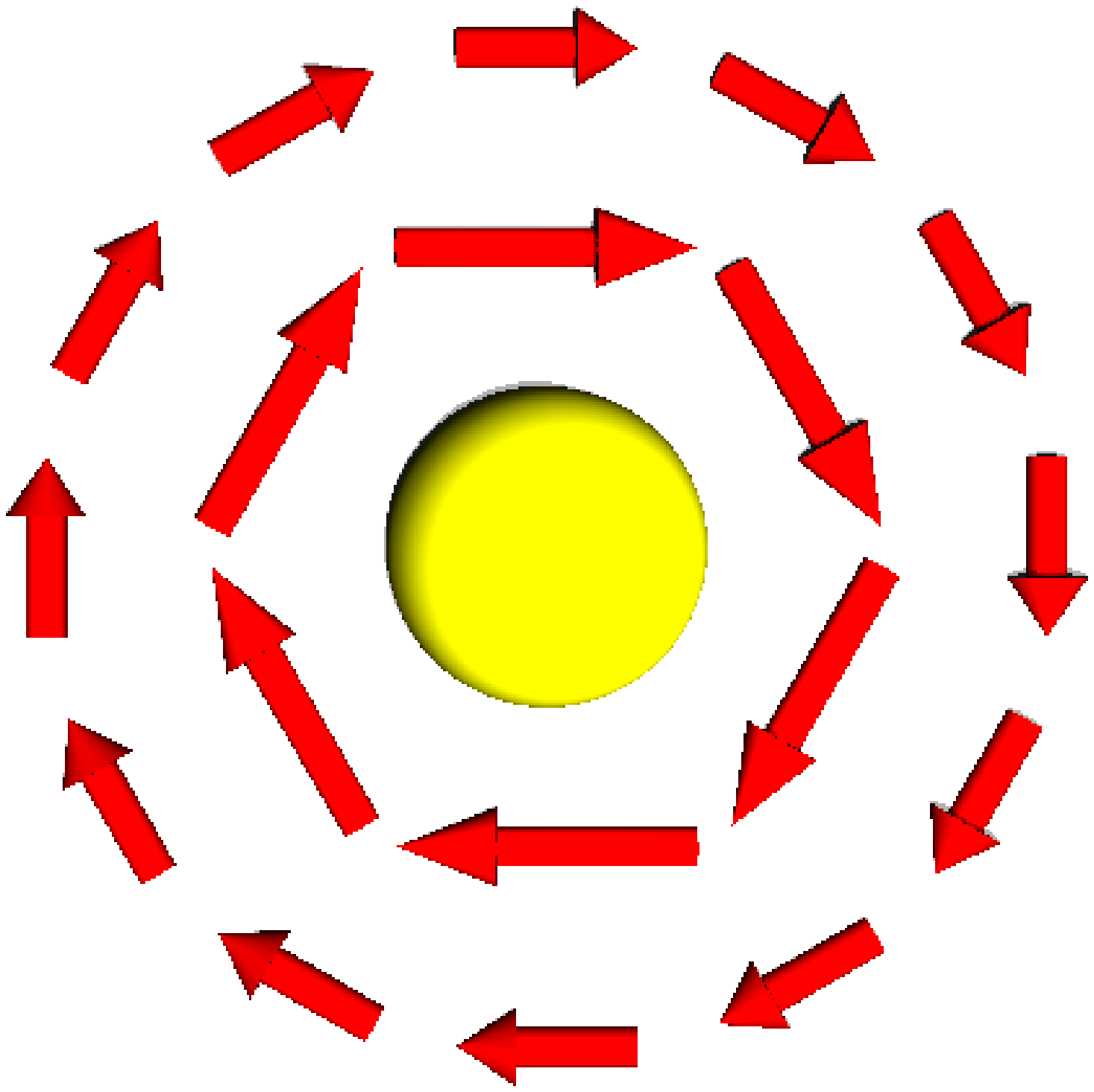}
  \end{minipage}
  \begin{minipage}{0.245\textwidth}
    \includegraphics[width=\textwidth]{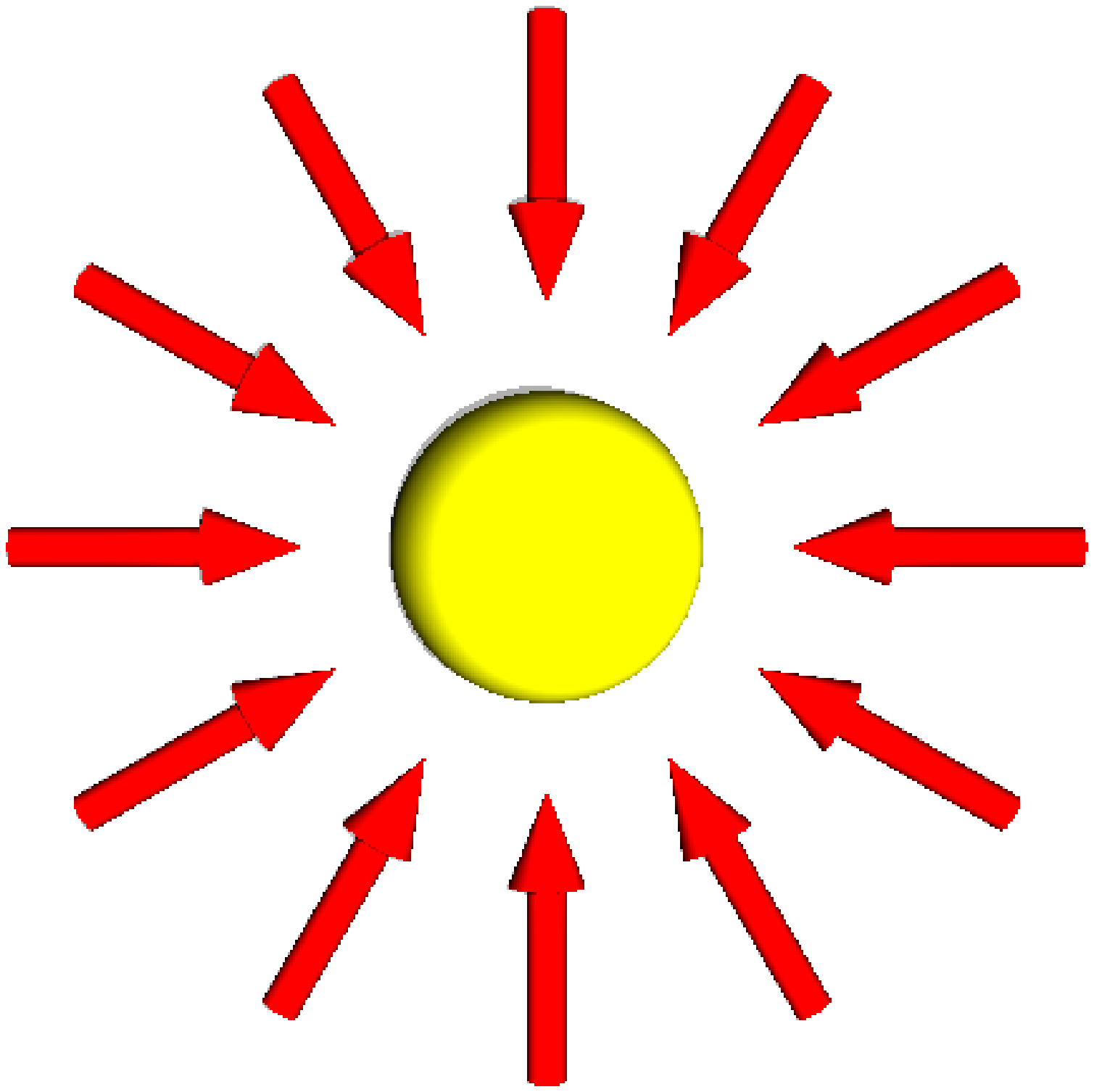}
  \end{minipage}
  \begin{minipage}{0.245\textwidth}
    \includegraphics[width=\textwidth]{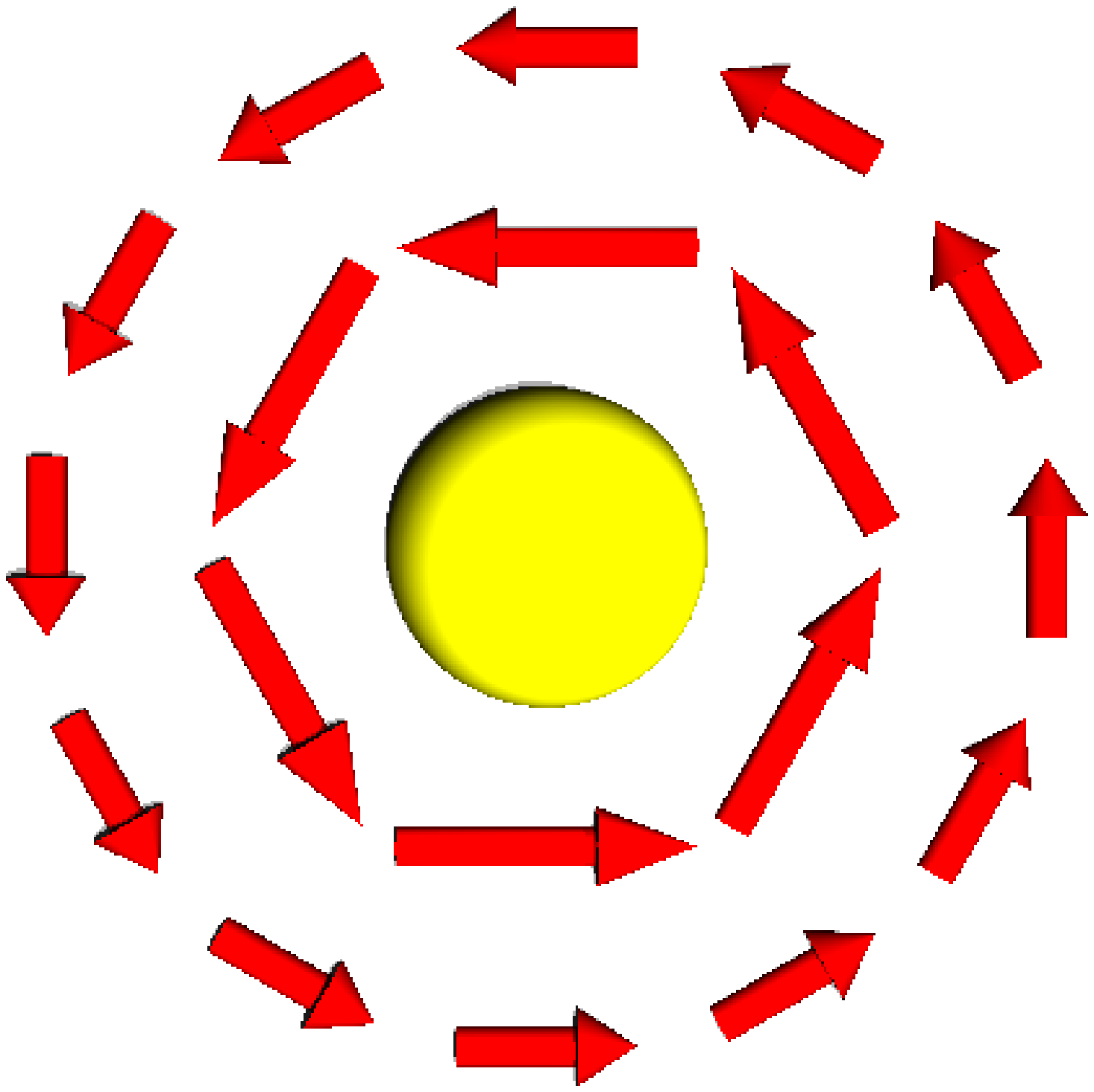}
  \end{minipage}
  \caption{
    Typical field configurations for topological defects in two
    dimensions.
    The shown field distributions (hedgehog, left vortex, 
    inverted hedgehog, and right vortex)
    can be transformed into one another
    by (internal) $SO(2)$ rotations, which preserve winding number.
    Hence, all four belong to the same homotopy class and have the
    same windung number $\mathfrak N = +1$.
  }
  \label{2Ddefects}
\end{figure}

\begin{figure}[hbt]
  \begin{center}
    \includegraphics[width=0.4\textwidth]{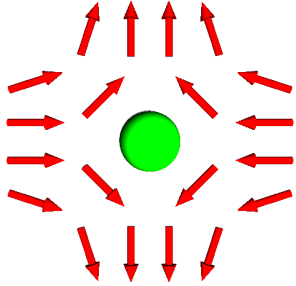}
  \end{center}
  \caption{Antidefect with $\mathfrak N = -1$ (``hyperbolic defect'')  
    which follows from
    any of the plots in Figure \ref{2Ddefects} after field inversion
    in one direction, i.e., after an improper rotation.
    \label{2Dantidefect}
  }
\end{figure}

To exemplify this, let us consider the simplest nontrivial case,
$N = 2$.
Parametrizing the field direction through a single angle,
$\hat n^1 = \cos\hat\theta$ and $\hat n^2 = \sin\hat\theta$, we
obtain
    \bea
    \hat{\mathfrak N} \, = \,
    \frac{1}{2\pi} \oint\limits_{\mathcal C}
    d\bm l \cdot\na\hat\theta \,,
    \label{2dwind}
    \ea
where $d\bm l$ is a tangent vector along the integration contour $\mathcal C$.
A rotation of the field direction
$\hat{\bm n}' = D(\phi) \hat{\bm n}$ by an angle $\phi$ will only add
a phase $\phi$ to the angle operator $\hat \theta$ and the winding
number \eqref{2dwind} is unaffected, see also Figure \ref{2Ddefects}.
Besides rotations, the group $O(N)$ also contains improper rotations,
i.e., combinations of a rotation with a reflection, which do not
conserve winding number.
%
%
This leads us, apart from the well-known monopole or spin-vortex
configurations depicted in Fig.\,\ref{2Ddefects}, to field
distributions with a nontrivial winding configuration, where 
the field points away in one direction and towards the defect core in
the other, see Fig.\,\ref{2Dantidefect}.


This can be generalized to higher dimensions.
For example, in three dimensions, any rotation of $\hat{\bm n}$ can be
parametrized by three Euler angles, i.e., as a sequence of three
two-dimensional rotations.
Obviously, inversion of an odd number of field direction, e.g.,
$\bm n \to -\bm n$, cannot be expressed in that way but corresponds to
improper rotations again so that the winding number picks up an
additional factor $-1$.
Hence, the typical $\mathfrak N = \pm1$ configurations have the field
pointing either away from ($+1$) or towards ($-1$) the defect core (in
all directions).
Other nontrivial field distributions, e.g., those with $\bm n$
pointing away in some and towards the core in all other directions,
can be obtained through rotations.
The strikingly different appearance of positive and negative defects
in two and three dimensions is typical for even and odd number of
dimensions $N$:
Field inversion, $\bm n \to -\bm n$, will leave the winding
number $\mathfrak N \to (-1)^N \mathfrak N$ invariant in even
dimensions while for $N$ odd, $\mathfrak N$ picks up a factor
$(-1)^N = -1$.
As we will see in an example below, this has implications
for the net defect number created during a quench to the
broken-symmetry phase.

\subsection{Winding number variance}

Since, for a fully $O(N)$ invariant initial state, 
the probabilities for creating positive and negative defects are
equal, the expectation value $\langle\hat{\mathfrak N}\rangle$ 
of the winding number operator \eqref{windop} is zero,
but the variance $\langle\hat{\mathfrak N}^2\rangle$ is
generally non-vanishing and can be used to quantify the
net number of defects created during the quench.
Due to the totally antisymmetric tensor $\varepsilon_{abc...}$ on
the right hand side of Eq.\ \eqref{windop}, each field component
$\hat n_a$ must appear exactly once in $\hat{\mathfrak N}$.
Hence, if different field components are uncorrelated,
cf.\ Eq.\ \eqref{corr}
     \bea
     \left\langle \hat{n}^a(\bm r) \hat{n}^b(\bm r')
     \right\rangle
     \, = \, f(\bm r - \bm r') \delta^{ab}
     \, = \, f(|\bm r - \bm r'|) \delta^{ab} \,,
     \label{2p}
     \ea
the expectation value $\langle\hat{\mathfrak N}^2\rangle$ must break
down into the product of $N$ two-point functions.
The first identity in \eqref{2p} holds when the system is homogenous,
while the second is valid when the system is isotropic as well.
This particular form of the two-point function
$\langle\hat n^a(\bm r)\hat n^b(\bm r)\rangle$ can be inferred from
symmetries:
Although the phase with broken $O(N)$ symmetry is neither homogeneous
nor isotropic, the expectation values still are, because we start from
an homogeneous and isotropic phase and any realization of the symmetry
breaking is equally likely.

With \eqref{2p}, there results the following expression
for the winding number variance
      \bea\fl 
        \langle\hat{\mathfrak N}^2(\mathcal M)\rangle  &= &
                \frac{N}{\|\mathcal S_{N-1}\|^2}
                \oint\limits_{\partial\mathcal M}
                        dS_\alpha \varepsilon^{\alpha\beta\gamma...}
                \oint\limits_{\partial\mathcal M}
                        dS'_{\alpha'} \varepsilon^{\alpha'\beta'\gamma'...}
        \nn\fl 
        && \times \left[ f(\partial_\beta \partial'_{\beta'}f)
                - (N-1) (\partial_\beta f)(\partial'_{\beta'}f) \right]
        (\partial_{\gamma}\partial'_{\gamma'} f)
        (\partial_\delta \partial'_{\delta'} f) ... \,,
        \label{N2}
      \ea
where we exploited the symmetry under exchange of index pairs
$(a,\alpha) \leftrightarrow (b,\beta)$, and also changed the order of
taking spatial derivatives and expectation values.
%
Due to the total antisymmetry of the integrand, this expression can be
regarded as a collection of surface integrals over $(N-1)$-forms (in  
the unprimed
indices while keeping the primed indices fixed and {\em vice versa}).
It becomes therefore possible to apply the generalized {Stokes} 
theorem,
$\int_{\mathcal M} d\omega = \int_{\partial \mathcal M} \omega$, and
rewrite \eqref{N2} as a volume integral over $N$-forms
      \bea\fl 
        \langle \hat{\mathfrak N}^2(\mathcal M) \rangle  =
                \frac{N^2}{\|\mathcal S_{N-1}\|^2}
                \varepsilon^{\alpha\beta\gamma...}
                \varepsilon^{\alpha'\beta'\gamma'...}
                \int\limits_{\mathcal M} d^Nr \, d^Nr'
                (\partial_\alpha\partial'_{\alpha'} f)
                (\partial_\beta\partial'_{\beta'} f)
                (\partial_\gamma\partial'_{\gamma'} f)...
        \,.
                  \label{N2vol}
      \ea
This expression is generally valid provided different field direction
are uncorrelated,
$\langle\hat n_a\hat n_b \rangle \propto \delta_{ab}$,
cf.\ \eqref{2p}.
The next step is less obvious, but again we can take advantage of the
inherent symmetries which enabled us to write the two-point function
\eqref{2p} through a function depending only on the separation
$L = |\bm r - \bm r'|$.
Using the identity
      \bea
        \partial_\alpha \partial'_{\alpha'} f =
                (\partial_\alpha\partial'_{\alpha'} L^2)
                \frac{\partial f}{\partial L^2}
                + (\partial_\alpha L^2) (\partial'_{\alpha'} L^2)
                \frac{\partial^2 f}{\partial L^4} \,,
      \ea
the derivatives under the integral can be evaluated.
In view of the total antisymmetry of the Levi-Civita symbols,
all terms containing more than two first derivatives of $L^2$, e.g.,
$(\partial_\alpha L^2)(\partial'_{\alpha'}L^2)(\partial_\beta L^2)
(\partial'_{\beta'}L^2)$ cancel after summation.
It follows
      \bea\fl 
        \langle \hat{\mathfrak N}^2 \rangle  =
                \frac{N^2}{\|\mathcal S_{N-1}\|^2}
                \varepsilon^{\alpha\beta\gamma...}
                \varepsilon^{\alpha'\beta'\gamma'...}
                \int d^Nr\, d^Nr'
                \left\{
                        (\partial_\alpha\partial'_{\alpha'}L^2)
                                \frac{\partial f}{\partial L^2}
                        + N (\partial_\alpha L^2)(\partial'_{\alpha'}L^2)
                                \frac{\partial^2 f}{\partial L^4}
                \right\}
        \nn \qquad \qquad \qquad \qquad \times
                (\partial_\beta\partial'_{\beta'} L^2)
                (\partial_\gamma\partial'_{\gamma'} L^2)...
                \left(\frac{\partial f}{\partial L^2} \right)^{N-1} \,.
      \ea
So far, we did not specify any particular coordinates, but in order to
simplify this expression, it is advantageous to adopt Cartesian
coordinates for which
$\partial_\alpha\partial'_{\alpha'}L^2 = - 2 \delta_{\alpha\alpha'}$.
After performing the summations over $\alpha$, $\alpha'$ etc.\ and
with $\partial f/\partial L^2 = (1/2L) \partial f/\partial L$, we
finally obtain
      \bea
        \langle \hat{\mathfrak N}^2(\mathcal M) \rangle =
                \frac{NN!}{\|\mathcal S_{N-1}\|^2}
                \int\limits_{\mathcal M} d^Nr \,d^Nr'
                \frac{1}{L^{N-1}} \frac{\partial}{\partial L}
                \left(-\frac{\partial f}{\partial L}\right)^N \,,
        \label{N2allg}
      \ea
which holds for arbitrary integration manifolds $\mathcal M$ and any
dimension $N$ provided the two-point correlator has the form \eqref{2p},
i.e., the correlator $\langle\hat n^a(\bm r)\hat n^b(\bm r')\rangle$
depends only on the separation between $\bm r$ and $\bm r'$.
If, however, homogenity or isotropy are broken but different directions
are still uncorrelated, the most general expression
for the winding number variance is given by Eq.\ \eqref{N2vol}.


\subsection{Spherical volumina}

The $2N$ integrals appearing in \eqref{N2allg} are, for arbitrary
volumina, generally difficult to evaluate directly,
but can be simplified drastically by exploiting symmetries of the
integration manifold $\mathcal M$.
In particular, we will demonstrate that for a sphere of radius $R$
-- an isotropic manifold defined through
$\mathcal M = \{\bm r: \bm r^2 < R^2\}$ --
Eq.\ \eqref{N2allg} reduces to a single integration over one
remaining variable.
To this end, let us instead of the winding number variance
$\langle\hat{\mathfrak N}^2\rangle$ consider its change with radius
$R$ of the integration volume.
Taking the derivative of \eqref{N2allg} with respect to $R$ and
using spherical coordinates, one of the radial integrations will
break down to yielding a factor $R^{N-1}$ from the integral measure
[note that the integrand of \eqref{N2allg} is symmetric in $\bm r$ and
$\bm r'$]. 
One of the angular integrations $d^{N-1}\Omega'$ can be performed
due to isotropy of the manifold $\mathcal M$
and gives the surface area $\|\mathcal S_{N-1}\|$ of the $N$ dimensional
unit sphere.
For the radial derivative of $\langle\hat{\mathfrak N}^2\rangle$, we get 
      \bea
        \frac{\partial \langle \hat{\mathfrak N}^2 \rangle}{\partial R}
        & = & 2 \frac{N N!}{\|\mathcal S_{N-1}\|} R^{N-1}
                \int\limits_0^R dr
                \int d^{N-1}\Omega\,
                \frac{J_N}{L^{N-1}}
                \frac{\partial}{\partial L}
                \left[-\frac{\partial f}{\partial L}\right]^N \,.
        \label{N2_sphere_step}
      \ea
The remaining volume integral $d^Nr = J_N\, dr d^{N-1}\Omega$ 
{with the Jacobian $J_N$} can be
tackled by a coordinate transformation
$\bm r \to \bm L = \bm r - \bm r'$.
Adopting spherical coordinates for $\bm L$, we see that the (radial)
measure $L^{N-1}$ just cancels the factor $1/L^{N-1}$ such that the
remaining $L$ dependence is just a partial derivative and the
modulus integral $dL$ can be performed easily.
Regarding the angular part, we exploit isotropy of $\mathcal M$ once
more and choose the angles $\phi_1...\phi_{N-1}$ such that the
separation $L = 2 R \sin\phi_1$ depends only on $\phi_1$.
After substituting $y = \sin\phi_1$ and performing the integrations
over the remaining angles, we then further get 
      \bea
        \frac{\partial \langle\hat{\mathfrak N}^2 \rangle}{\partial R}
        & = & 2 N N! \frac{\|\mathcal S_{N-2}\|}{\|\mathcal S_{N-1}\|}
                R^{N-1}
        \int\limits_0^1 dy \sqrt{1 - y^2}^{N-3}
                \left[-\frac{\partial f}{\partial L}(2Ry)\right]^N \,.
        \label{dN2dR}
      \ea
Using the identity $\sqrt{1-y^2}^{N-3} = - (1/N-1)y^N
\partial(\sqrt{1-y^2}^{N-1}/y^{N-1})/\partial y$, Eq.\ \eqref{dN2dR}
can be rewritten and, after integration by parts,
the change of the winding number variance reads
      \bea \fl 
        \frac{\partial\langle \hat{\mathfrak N}^2\rangle}{\partial R}
                 &=& 2 \frac{NN!}{N-1}
                \frac{\|\mathcal S_{N-2}\|}{\|\mathcal S_{N-1}\|}
                \int\limits_0^1 dy \frac{\sqrt{1-y^2}^{N-1}}{y^{N-1}}
                \frac{1}{R}
                \frac{\partial}{\partial y} (Ry)^N
                \left[-\frac{\partial f}{\partial R}(2Ry)\right]^N .
      \ea
In the factors right of the derivative, the integration variable $y$
appears only in the combination $Ry$.
Hence, using the chain rule, $\partial/\partial y$ can be replaced
by $(R/y) \partial/\partial R$ and it follows
      \bea  
        \frac{\partial\langle\hat{\mathfrak N}^2\rangle}{\partial R}
         =  \frac{NN!}{N-1}\frac{\|\mathcal S_{N-2}\|}{\|\mathcal S_{N-1}\|}
                \frac{\partial}{\partial R} R^N
                \int\limits_0^1 dy \sqrt{1-y^2}^{N-1}
                \left[-\frac{\partial f}{\partial L}(2Ry)\right]^N \,.
      \ea
Both sides of this equation are partial derivatives with respect to the
manifold radius $R$ and can be formally integrated, where the constant
of integration follows by requiring
$\langle\hat{\mathfrak N}^2(R = 0)\rangle = 0$.
Finally with $y = \sin(\theta)$, we can simplify the result to the more compact form
      \bea
        \langle\hat{\mathfrak N}^2(R)\rangle
        & = & \frac{N!}{\pi} R^N \int\limits_0^{\pi/2}
                d\theta \left[
                        -\cos \theta
                        \frac{\partial f}{\partial L}
                                (2R\sin\theta)
                \right]^N \,,
                  \label{Nsphere}
      \ea
i.e., the winding number variance can be determined through a
single angular integration.
For short-ranged correlations, where $\partial f/\partial L$ 
falls off fast enough at large distances $L$ 
\bea
\frac{\partial f}{\partial L}\leq\ord(L^{-\xi})\;:\,\xi>\frac{1}{N}
\,,
\label{shortrange}
\ea
%
a surface scaling \cite{PRL,PRD}
\bea
\langle\hat{\mathfrak N}^2(R)\rangle \propto R^{N-1}
\label{Nshortrange}
\ea
follows for large $R$:
For sufficiently large $R$, 
the main contribution to \eqref{Nsphere} arises from small $\theta$.
By formally extending the integration to infinity and approximating
$\cos\theta \approx 1$ and $\sin\theta \approx \theta$, one can see that
the integral is proportional to $1/R$ and we obtain scaling of
the winding number variance with the surface area of the enclosed
volume.
Note that expression \eqref{Nsphere} and thus also the surface scaling
for short-ranged correlations \eqref{shortrange} 
hold in {\em any} dimension $N$ as long as the direction 
correlator takes the form \eqref{2p}.
In particular, the expression for $\langle\hat{\mathfrak N}^2\rangle$
derived in \cite{PRL} will be reproduced for $N = 2$.


We stress that our calculation based on very general principles 
predicts a {\em surface scaling law} for the net defect number variance. 
This surface law should be contrasted with the {\em volume scaling law} 
obtained from another frequently used and phenomenologically motivated approach
\cite{Zurek}.
If one would randomly distribute hedgehogs and antihedgehogs in a given 
$N$-dimensional space with the same fixed densities, the total number of 
defects (hedgehogs plus antihedgehogs) inside a given volume would scale 
with the volume itself. 
Their difference (i.e., hedgehogs minus antihedgehogs), however, would 
scale with the square root of that volume (corresponding to a one-dimensional 
random walk of the winding number) and thus the net defect number variance 
scales with the volume. 
%

As far as we are aware, the rather generic result obtained here, together 
with our previous results in \cite{PRL,PRD}, represent the first proof by direct
calculation that (the quantum version of) the Kibble-Zurek mechanism leads to a surface law.


\subsection{General convex volumina}

The surface scaling for short-ranged interactions can be interpreted
as a variety of a "confined" phase, 
where bound defect-antidefect pairs, typically
separated by a correlation length, occur.
Pairs entirely within or outside the manifold $\mathcal M$ will not
contribute to the net winding number and thus also not to
$\langle\hat{\mathfrak N}^2\rangle$.
Only those pairs, where one of the partners is inside and the other
outside will contribute.
Since it is not clear whether the defect or the antidefect is contained
within $\mathcal M$, calculation of the winding number would be similar
to a random walk with number of steps proportional to surface area.
From this simple picture, it can be anticipated that the surface scaling
is not restricted to spherical manifolds but might apply in any situation 
where the surface of the manifold is sufficiently "smooth".

In the following, we will thus consider arbitrary convex manifolds and show
that the surface area scaling of the winding number variance is retained.
We parametrize the convex manifold in spherical coordinates
      \bea
      \mathcal M = \left\{ \bm r: \bm r^2 <
      R^2 a^2(\vartheta_1,...\vartheta_{N-1}) \equiv
      R^2 a^2(\bm \Omega) \right\}
      \nn
      {\rm with} \quad
      s \bm r_1 + (1-s)\bm r_2 \in V \quad \forall\;\bm r_1, \bm r_2 \in V, \;
      s \in [0,1]
      \label{manifold}
      \ea
through a scale factor $R$ and a shape function $a(\bm \Omega) = \ord(1)$,
similar to a sphere, where we would have $a(\bm \Omega) \equiv 1$.
Obviously, the surface area $\partial\mathcal M \propto R^{N-1}$ so that
we need to show $\langle\hat{\mathfrak N}^2\rangle \propto R^{N-1}$ as well.
For convex manifolds, the volume element in expression \eqref{N2allg} for
the winding number variance reads in spherical coordinates
      \bea
      \int d^Nr = \int d^{N-1}\Omega J_N(\bm \Omega)
      \int \limits_0^{Ra(\bm \Omega)} dr r^{N-1}
      \,,
      \ea
where the Jacobian $J_N(\bm \Omega)$ is taken at $r \equiv 1$.
As for spherical volumina, cf.\ \eqref{N2_sphere_step}, the scale factor $R$
appears only in the upper bound of the radial integration such that we
can use the same trick as before and calculate
$\partial\langle\hat{\mathfrak N}^2\rangle/\partial R$ instead of the variance
itself
      \bea \fl 
      \frac{\partial\langle\hat{\mathfrak N^2}\rangle}{\partial R}
       =  2 \frac{NN!}{\|\mathcal S_{N-1}\|^2} R^{N-1}
      \int d^{N-1}\Omega'J_{N}' a^N(\bm \Omega')
      \int d^Nr \frac{1}{L^{N-1}}\frac{\partial}{\partial L}
      \left(- \frac{\partial f}{\partial L}\right)^N \,.
      \ea
However, in contrast to the spherical case, we cannot simply perform
the angular $d^{N-1}\Omega$ integration because of broken isotropy (of
the manifold):
The separation $L = |\bm r - \bm r'|$ generally depends on both
azimuthal positions $\bm \Omega$ as well as $\bm \Omega'$ and not only
on their relative angles.
The coordinate transformation $\bm r \to \bm L = \bm r - \bm r'$ and
subsequent integration in spherical coordinates yields
      \bea\fl 
      \frac{\partial \langle\hat{\mathfrak N}^2\rangle}{\partial R}
      =  2 \frac{NN!}{\|\mathcal S_{N-1}\|^2} R^{N-1}
      \int d^{N-1}\Omega'd\tilde\Omega J_N(\bm\Omega')
      J_N(\tilde{\bm\Omega})
      a^N(\Omega')
      \left( -\frac{\partial f}{\partial L}
      \left[L_{\rm max}(\tilde{\bm\Omega},\bm\Omega')\right]\right)
      \,, \nn
      \label{N2byR}
      \ea
where $L_{\rm max}$ is the distance from a point $\bm\Omega'$ on the
surface $\partial\mathcal M$ to the opposite boundary of $\mathcal M$
in direction $\tilde{\bm\Omega}$ and is determined through
$L_{\rm max} =
|\bm r_{\rm max}[\bm \Omega(\tilde{\bm \Omega})] -
\bm r'_{\rm max}(\bm \Omega')|$.
[At this point, it should be mentioned that one needs to be careful
regarding the azimuthal positions:
$\bm\Omega'$ and $\bm \Omega$ are those of $\bm r'$
and $\bm r$ as seen from the predefined centre of the manifold,
cf.\ \eqref{manifold},
whereas $\tilde{\bm \Omega}$ is the azimuthal position of $\bm r$ as
seen from $\bm r'$.]
A scaling of the winding number variance with surface area is obtained
if the integral in \eqref{N2byR} is proportional to $1/R$ for
sufficiently large $R$ -- which can indeed be expected for short-ranged
correlations \eqref{shortrange} 
because $\bm r_{\rm max}$ and $\bm r'_{\max}$ both scale with $R$ and thus
$L_{\rm max} = R \Lambda_{\rm max}(\tilde{\bm \Omega},\bm \Omega') \propto R$.

We evaluate the scaling behaviour of \eqref{N2byR} by choosing
suitable coordinates for the integration:
In view of the convexity of $\mathcal M$, the angles
$\tilde{\bm\Omega}$ can be chosen such that for each $\bm \Omega'$
there exists at least one angle $\phi_1$ of $\tilde{\bm \Omega}$ for
which the separation $L_{\rm max} = R \Lambda_{\rm max}$ changes from
zero to order $R$ while keeping all other angles $\phi_2$ to
$\phi_{N-1}$ fixed at arbitrary values.
The $\phi_1$ integral appearing in \eqref{N2byR} will be performed
first, it reads
      \bea
      \int\limits_0^{\pi} d\phi_1\, \sin^{N-2}\phi_1
      \left[ -\frac{\partial f}{\partial L}
        [ R\Lambda_{\rm max}(\phi_1) ]\right]^N .
      \label{convex_phi}
      \ea
Obviously, the integrand vanishes for all $\phi_1$ with
$\Lambda_{\rm max}(\phi_1) = 0$.
It becomes thus possible to change the domain of integration to
$[\phi_-, \phi_+] \subset (0,\pi)$ where $\Lambda_{\rm max}(\phi) > 0$
for all $\phi \in (\phi_-,\phi_+)$ and
$\Lambda_{\rm max}(\phi) = 0 \, \forall\, \phi \notin (\phi_-,\phi_+)$.
 From convexity of the integration manifold $\mathcal M$ also follows
that the maximal separation $\Lambda_{\rm max}$ cannot have (local)
minima in $(\phi_-,\phi_+)$, i.e, $\Lambda_{\rm max}$ must be
monotonically growing in the vicinity of $\phi_-$ and decreasing near
$\phi_+$.
Furthermore, by demanding short-ranged correlations, \eqref{shortrange},
$\partial f/\partial L$ becomes arbitrarily small for large separations.
(See, by contrast, the case of long-range correlations we discuss in
the following section).
This means that, for sufficiently large $R$, the main contributions to
integral \eqref{convex_phi} stems from regions where $\Lambda_{\rm max}$
is small but non-zero, i.e., near $\phi_-$ and $\phi_+$, and we
can approximate \eqref{convex_phi}
     \bea\fl 
        \int\limits_0^{\pi} d\phi_1\, \sin^{N-2}\phi_1
        \left[ -\frac{\partial f}{\partial L}
        [ R\Lambda_{\rm max}(\phi_1) ]\right]^N
        & \approx & \sum_{a=\pm} \,
        \int\limits_{\phi_a -\epsilon}^{\phi_a +\epsilon}
        d\phi_1\, \sin^{N-2}\phi_1
        \left(-\frac{\partial f}{\partial L} \right)^N
        \label{phi_convex_2}
     \ea
with small $\epsilon \ll 1$.
For all angles $\phi \in (\phi_- +\epsilon,\phi_+ -\epsilon)$, 
the distance 
$\Lambda_{\rm max} > 0$ but the integrand is negligibly 
small due to short-range correlations \eqref{shortrange}.
In the integrals on the right hand side of \eqref{phi_convex_2}, we
can assume that $\Lambda_{\rm max}$ is strictly monotonic and analytic
such that $\Lambda_{\rm max}(\phi_1)$ is bijective in
$(\phi_-, \phi_-+ \epsilon)$ and similarly in $(\phi_+ -\epsilon,\phi_+)$.
It becomes therefore possible to change the integration variable from
$\phi_1$ to $\Lambda_{\rm max}$ and we finally obtain for
the $\phi_1$ integral \eqref{convex_phi}
     \bea\fl 
        \int\limits_0^{\pi} d\phi_1\, \sin^{N-2}\phi_1
                \left[ -\frac{\partial f}{\partial L}
                [ R\Lambda_{\rm max}(\phi_1) ]\right]^N
        \label{convex_phi_3}\\
        \phantom{+}
        \approx \sum_{a=\pm} \frac{\sin^{N-1}\phi_a}{R}
        \int\limits_0^\infty
        d(R\Lambda_{\rm max})
        \left(
        \frac{\partial \Lambda_{\rm max}}{\partial \phi_1}(\phi_a)
        \right)^{-1}
        \left(-\frac{\partial f}{\partial L}(R\Lambda_{\rm max}) \right)^N
     \nonumber\ea
where we have extended the integration to infinity because, due to
short-range correlations, the integrand becomes negligibly small
for sufficiently large $R$.
In this limit, the integral yields a constant (independent of system
size $R$ but of course still depending on all other angles
$\phi_2$...$\phi_{N-1}$ as well as $\bm \Omega'$) and the scale factor
$R$ appears only as prefactor $1/R$.
We are thus led to the scaling of \eqref{N2byR} with $R^{N-2}$
for large $R$ and thus, after integration, obtain for the
winding number variance
      \bea
      \langle \hat{\mathfrak N}^2 (R)\rangle \propto R^{N-1},
      \label{scaleconvex}
      \ea
i.e., the desired surface scaling result.
It should be noted that the $R^{N-2}$ dependence of
$\partial\langle\hat{\mathfrak N}^2\rangle/\partial R$ derived in this
section is only valid for sufficiently large $R$.
When integrating
$\partial\langle\hat{\mathfrak N}^2\rangle/\partial R$ to obtain the
scaling behavior of the winding number variance, regions with small
$R$, where $\partial\langle\hat{\mathfrak N}^2\rangle/\partial R$
might behave differently, are also included, and potentially
contribute subleading orders to the winding number variance, which
scales with surface area for large $R$.

\section{Deviations from the Surface Law}

We have obtained the surface scaling \eqref{Nsphere} for spherical
volumes in high dimensions, and  generalized this result to arbitrary
convex volumes \eqref{scaleconvex}.
The salient ingredients of the derivation have been 
%
%
%
%
the short-range nature of correlations, Eq.\,\eqref{Gaussian}, 
and the symmetries of the correlation function \eqref{2p}.
By relaxing either or both of these conditions, 
and depending on the precise functional form of the two-point correlator,
it is then possible, as we will now show, to obtain
various interesting scaling behaviors with system size.
%



Let us consider broken spatial isotropy of the correlation function
\eqref{2p}, e.g., through an externally imposed lattice.
(Note that $\langle \hat n_a \hat n_b \rangle \propto \delta_{ab}$ is
still isotropic in the field directions.)
Explicit examples for periodic and short-ranged correlations will be
discussed in subsections \ref{sec_periodic} and \ref{sec_gauss}.
%
%
To remain general, we {assume} that in \eqref{2p} only the first 
equality holds, and that the (real space part of) the correlator separates 
into the product
      \bea
      f(\bm r - \bm r') \, = \,
	\prod_{\alpha=1}^N g_\alpha(r_\alpha - r_\alpha') \,,
      \label{sepf}
      \ea
where the correlations $g_\alpha$ in different directions can in general have a
different functional behavior.
We insert this expression into \eqref{N2vol}, and consider the
volume enclosed by a $N$-dimensional box with size $L_1\times L_2  
\times L_3...$.
The second-order partial derivative of the correlator \eqref{sepf} gives
      \bea
      \partial_\alpha \partial_{\alpha'}' f
      \, = \, - f \left[
      \frac{ g_\alpha'}{g_\alpha} \frac{g_{\alpha'}'}{g_{\alpha'}}
      + \left( \frac{g_\alpha''}{g_\alpha}
      - \frac{g_\alpha'}{g_\alpha}\frac{g_\alpha'}{g_\alpha}
      \right) \delta_{\alpha\alpha'}\right]
	\label{ddsepf}
      \ea
where a prime denotes the derivative, e.g.,
$g_\alpha' = \partial g_\alpha(x)/\partial x$.
Inserting the above expression into \eqref{N2vol}, we see that any terms
involving products $(g'_{\alpha}/g_\alpha) (g'_{\alpha'}/g_{\alpha'})
(g'_{\beta}/g_\beta) (g'_{\beta'}/g_{\beta'})$ etc.\ cancel
during summation due to antisymmetry of the $\varepsilon$ pseudotensors
and we have
      \bea
\fl	
\left\langle\hat{\mathfrak N}^2\right\rangle
	\, = \,
	(-1)^N \frac{N^2}{\|\mathcal S_{N-1}\|^2}
	|\varepsilon^{\alpha\beta\gamma...}|^2
	\int d^Nr \int d^Nr' f^N
	\left(\frac{g_\alpha''}{g_\alpha}
	+(N-1)\frac{g_\alpha'}{g_\alpha}\frac{g_\alpha'}{g_\alpha} \right)
	\nn	\times
	\left(
	\frac{g_\beta''}{g_\beta} -
	\frac{g_\beta'}{g_\beta}\frac{g_\beta'}{g_\beta}
	\right)
	\left(
	\frac{g_\gamma''}{g_\gamma} -
	\frac{g_\gamma'}{g_\gamma}\frac{g_\gamma'}{g_\gamma}
	\right)... \,.
      \ea
This correlation function can be evaluated further by accounting for
the broken isotropy of $f$ and regarding the integral over each
each spatial direction independently.
It follows
    \bea
\fl	
\left\langle\hat{\mathfrak N}^2\right\rangle
	\, = \,
	 \frac{2(-1)^{N-1}NN!}{\|\mathcal S_{N-1}\|^2}
	\sum_\alpha [g_\alpha^N(0) - g_\alpha^N(L_\alpha)]
	\prod_{\beta\neq \alpha}
	\int\limits_0^{L_\beta} dr_\beta
	\!
	\int\limits_0^{L_\beta} dr_\beta'
	\left[g_\beta^{N-1} g_\beta''
	- g_\beta^{N-2} g_\beta' g_\beta' \right] .
	\nn\label{brokenisocorr} 
    \ea
So far, this result is general.
We now are going to
illustrate it with a few concrete examples for the correlation function.

\subsection{Periodic Order in the Direction Operator Correlations}
\label{sec_periodic}

{As an example for a nonlocalized correlation function,}
we consider a purely periodic functional behavior,
%
which can, for example, be derived within the following simple model. 
Using the ansatz for the fluctuating order-parameter 
field
	\bea
	\phi_a(\f{r})=\prod_b\cos(kx_b+\varphi_{ab})
	\ea
and averaging over $N \times N$ independent fluctuating angles
$\varphi_{ab}\in[0,2\pi]$, one obtains vanishing expectation
value
%
$	\langle\phi_a(\f{r})\rangle=
	\prod_{b}
	\frac{1}{2\pi}\int\limits_0^{2\pi}d\varphi_{ab}\,
	\cos(kx_b+\varphi_{ab})=0
$. 
The two-point correlations are then calculated to be
	\bea
	\langle\phi_a(\f{r})\phi_b(\f{r})\rangle=
	\frac12\delta_{ab}\prod_{c}\cos(k[x_c-x_c']),
	\label{fcos}
	\ea
and explicitly break the isotropy of the system by imposing periodicity
along all $N$ spatial axes.

The integrals in \eqref{brokenisocorr} now grow with system size 
$L_\alpha$ for an even number of dimensions (the integrand is 
strictly positive for even $N$), 
whereas the first term in the sum,
$g_\alpha^N(0) - g_\alpha^N(L_\alpha) = 1 - \cos^N(k L_\alpha)$, is
a periodic function in the system size $L_\alpha$.
Thus the correlation function \eqref{fcos} implies a scaling of the
winding number variance (for even $N$) with the square of the surface
area of integration volume $V = L_1 \times L_2 \times L_3...$
    \bea
	\left\langle\hat{\mathfrak N}^2\right\rangle_{{\rm even}\;{N}}
	\, \propto \,
	\sum_\alpha [1 - \cos^N(kL_\alpha)]\frac{V^2}{L_\alpha^2} \,. 
\label{periodic}
    \ea
The factors $1 - \cos^N(kL_\alpha)$, however, will still lead to
vanishing winding number variance if the size $L_\alpha = n \pi/k$
matches the periodicity of the correlations in {\em all} spatial directions.
For an odd number of dimensions, on the other hand, the integrands
of \eqref{brokenisocorr} oscillate periodically around zero and
no scaling of $\langle\hat{\mathfrak N}^2\rangle$ with system size
is obtained, $\langle\hat{\mathfrak N}^2\rangle=\ord(V^0)$.
%

\vspace*{1em}
\begin{figure}[!bh]
\bc
 \includegraphics[angle=0,width=0.5\textwidth]{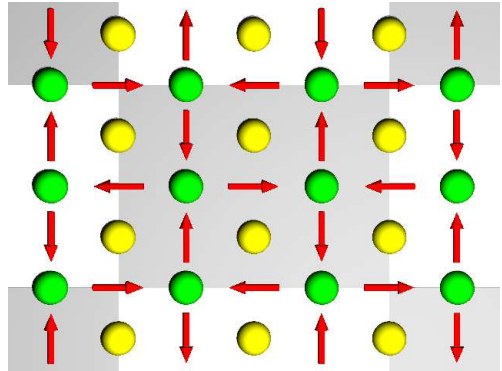}
\caption{\label{lattice} Distribution of topological defects 
  in the presence of periodic order in two dimensions.
  The unit cells of the lattice are depicted as gray and white areas,
  while the defect cores are depicted as yellow for ${\mathfrak N}=+1$ 
  and green for ${\mathfrak N}=-1$, cf.\ Figs.\ \ref{2Ddefects} and
  \ref{2Dantidefect}.
}
\ec
\end{figure}

This entirely different scaling -- surface squared with
oscillating prefactor for even dimensions versus essentially no
net defects in odd dimensions -- follows directly from the behavior
of the winding number under field inversion described in
Sec.\ \ref{SecDefects} (see also Figs.\,\ref{2Ddefects} and
\ref{2Dantidefect}):
Due to the strict periodicity of the correlation function
\eqref{fcos}, the field orientation must be inverted after half a unit
cell $\bm n(\bm r) = - \bm n(\bm r + \bm e_a \pi/k )$ in every
direction $\bm e_a$, see Fig.\,\ref{lattice}.
Hence, for each defect with topological charge $\mathfrak N = +1$,
there sits another defect with charge $\mathfrak N = (-1)^N$, 
a distance $\pi/k$ apart in every lattice direction.
This means for even dimensions, one has alternating layers of defects
and antidefects, and thus the surface squared scaling with oscillating
prefactor results, whereas in odd dimensions, each layer contains positive and
negative defects, so that the total winding number essentially
vanishes for a box-like integration manifold.
(However, if the box-shaped integration manifold were tilted with
respect to the lattice, a non-vanishing net defect number could occur
in odd dimensions as well.)


\subsection{Localized (Gaussian) Order}
\label{sec_gauss}

As a second example, we take short-ranged correlations, e.g., a
Gaussian correlator similar to \eqref{Gaussian} derived for
spatial isotropy in large $N$.
Now, each of the double integrals in \eqref{brokenisocorr}
scales with $L_\beta$, such that the winding number variance
is proportional to the surface area of the enclosed volume
    \bea
	\left\langle\hat{\mathfrak N}^2\right\rangle
	\propto
	\sum_\alpha
	\prod_{\beta\neq\alpha} L_\beta
	\int\limits_0^\infty dL
	\left[ g_\beta^{N-1}(L) g_\beta''(L)
	- g_\beta^{N-2} g_\beta'(L) g_\beta'(L) \right] \,.
    \ea
The above relation implies that the surface scaling found for spatially isotropic short-range 
correlations, cf.\ Eq.\ \eqref{scaleconvex}, 
 is shown to be preserved also for the case of broken spatial isotropy. This 
 should be contrasted with the result of the previous subsection, where it 
 was shown that for {\em long-range} periodic order, no scaling proportional
 to the surface area is obtained.

\section{Conclusion}

Starting from a rather general $O(N)$ symmetric action, 
we derived a quadratic effective action for the field
fluctuations, and studied the instability of modes during a
phase transition from the initial $O(N)$ symmetric (``paramagnetic'')  
phase to a broken symmetry phase in which ``ferromagnetic'' order  
develops. 
Since we are interested only in the exponentially growing modes, which
dominate the development of the domains and thus also of topological
defects, we introduced a systematic averaging procedure for the fluctuating field 
operator directions.
We have demonstrated that the direction field correlation function can 
be explicitly evaluated.
Our results are very general:
The initial state (such as temperature, dispersion relation) enters the 
coefficients $B_k$ and $C_k$ in (\ref{phi_quench}) only.
Thus, while the initial state affects the field correlator 
$\langle\hat\phi_a(t,\f{r})\hat\phi_b(t,\f{r'})\rangle$,
it leaves the direction $\f{n}$-correlator (\ref{hatnfull}) invariant. 
The external conditions after the transition determine the final 
dispersion relation which yields the value of $k_*$.
Apart from a simple rescaling of this value, our main results for the 
homogeneous and isotropic case (surface scaling law etc.) 
thus do not depend on the initial and the final state. 

%
%
%


This correlator of the field direction then directly determines the
statistics of the net defect number created by the quench.
A general expression for the winding number variance was derived and
was shown to scale with the surface area of the enclosed convex volume 
in any (large) dimension $N$.
Such a surface area scaling follows from general arguments based 
on the assumption of short-ranged correlations together with the 
internal $O(N)$ and spatial symmetries (homogenity and isotropy).
For large enough samples, it should be possible to verify this 
prediction experimentally.
The detection method depends on the specific experimental setup: 
In Bose-Einstein condensates, the topological defects are phase or 
spin vortices (in two spatial dimensions) which can be imaged 
either by absorption (resolving the core as a hole in the condensate \cite{Weiler}),  
or 
using polarization-dependent phase-contrast imaging \cite{Higbie}.
In superfluid $^3\!$He, vortices can be detected with NMR techniques \cite{Finne}.
Finally, in conventional condensed matter 
systems with magnetic order like 
ferromagnets, topological defects can be detected for example with Lorentz transmission
electron microscopy \cite{Uchida}. 


If the correlations are not short-ranged, the scaling law can be very different:
E.g., for purely periodic (i.e., long-ranged) correlations 
on a lattice, the scaling law strongly depends on the parity, 
i.e., whether the number of dimensions $N$ is even or odd.
In even dimensions, the winding number variance basically scales with
surface area squared with oscillating prefactor. 
By contrast, for odd $N$, the net defect number inside a box of size
$V = L_1 \times L_2 \times L_3...$ effectively vanishes.
This entirely different behavior can be explained by the strict
periodicity of the field directions together with properties of the
winding number operator, i.e., how positive and negative defects are
aligned on a lattice:
When inverting the field direction, $\bm n \to - \bm n$, the winding
number $\mathfrak N$ acquires a prefactor $(-1)^N$.
Hence, configurations where the field points towards to and those
where it points away from the defect core must lie in the same
homotopy class ($\mathfrak N = +1$) for even dimensions and in
different classes ($\mathfrak N = \pm 1$) in odd dimensions.
The strict periodicity of the correlation function, on the other hand,
implies a field inversion after half a period (along any of the
lattice directions).
Thus, one has alternating layers of positive and negative defects for 
even dimensions, while for odd dimensions positive and negative 
defects appear in each layer.

Further possible directions of research include the extension of our
results for the winding number variance to finite, experimentally
accessible values of $N$.
We stress in this context that the replacement of the quantum 
normalization factor $\hat Z$ introduced through the time-averaging of the direction vector, 
by its classical expectation value in the correlator \eqref{hatnfull}, remains
the only finite $N$ approximation made in our analysis. This statement holds 
under the condition that we have a system for which the effective action for 
the field remains still quadratic for finite $N$ (like in a spinor Bose-Einstein condensate)
leading to a dispersion relation with a dominant wavenumber $k_*>0$.
The finite $N$ extension can then be achieved by evaluating
Eq.\,\eqref{hatnfull} to the next order(s) in $1/N$, i.e., by
calculating the corrections to the operator-valued normalization
factor \eqref{n_def} entering the average quantum direction vector in
\eqref{average_n}.

\section*{Acknowledgements}
M.\,U. acknowledges support by the
Alexander von Humboldt Foundation and NSERC of Canada,
R.\,S. by the DFG (SCHU~1557/1-3, SFB-TR12), and
U.\,R.\,F. by the DFG (FI 690/3-1), 
Seoul National University, 
and the 
Basic Science Research Program of the National Research Foundation
 of Korea (NRF), Grant No. 2010-0013103. %


\end{document}